\documentclass[preprint,12pt]{elsarticle}

\usepackage{amssymb}
\usepackage{amsmath}

\journal{Physica E}

\begin{document}

\begin{frontmatter}



\title{Full counting statistics of Majorana interferometers}


\author{Gr\'egory Str\"ubi}
\address{Department of Physics, University of Basel,
  Klingelbergstrasse 82, CH-4056 Basel, Switzerland}
\author{Wolfgang Belzig}
\address{
Department of Physics, University of Konstanz, D-78457 Konstanz, Germany}
\author{Thomas L. Schmidt}
\address{Physics and Materials Science Research Unit, University of Luxembourg, L-1511~Luxembourg}
\author{Christoph Bruder}
\address{Department of Physics, University of Basel,
  Klingelbergstrasse 82, CH-4056 Basel, Switzerland}
\ead{christoph.bruder@unibas.ch}

\begin{abstract}
  We study the full counting statistics of interferometers for chiral
  Majorana fermions with two incoming and two outgoing Dirac fermion
  channels. In the absence of interactions, the FCS can be obtained
  from the $4\times4$ scattering matrix $S$ that relates the outgoing
  Dirac fermions to the incoming Dirac fermions.  After presenting
  explicit expressions for the higher-order current correlations for a
  modified Hanbury Brown-Twiss interferometer, we note that the
  cumulant-generating function can be interpreted such that
  unit-charge transfer processes correspond to two independent
  half-charge transfer processes, or alternatively, to two independent
  electron-hole conversion processes.  By a combination of analytical
  and numerical approaches, we verify that this factorization property
  holds for a general $SO(4)$ scattering matrix, i.e. for a general
  interferometer geometry.
\end{abstract}


\end{frontmatter}


\section{Introduction}
In a seminal work B\"uttiker pointed out the power of investigating
non-local current correlations in mesoscopic conductors to detect
particle exchange effects~\cite{buttiker1992}.  This work extended
the scope of previous studies of shot noise in two-terminal conductors
\cite{khlus,lesovik1989,buttiker1990,martin1992} by showing that
fundamental quantum mechanical effects can play a decisive role in
electronic transport properties.

Later on, B\"uttiker demonstrated on completely general grounds that
cross-correlations of bosonic and fermionic free particles are
fundamentally different~\cite{buttiker1992PRB}. Whereas for bosons the
sign is determined by the competition between positively contributing
bunching effects and negatively contributing partitioning effects,
free fermions always exhibit overall negative cross-correlations due
to their antibunching property.  Thus, the observation of positive
cross-correlations for fermions requires some nontrivial interaction
between them. A first example was provided by a superconducting source
injecting currents into two normal leads
\cite{martin1996,anantram1996}. Surprisingly, the effect persists even
for many-channel conductors
\cite{boerlin2002,samuelsson:02}. Furthermore, strong Coulomb
interactions can also impose positive cross-correlations in
multi-terminal quantum dot systems \cite{cottet2004,cottet2004EPL}.
An analysis of the full counting statistics reveals a dynamical
bunching effect as the origin of the positive cross-correlations
\cite{boerlin2002,belzig2005}.

Because of their fascinating properties and potential applications in
topological quantum information~\cite{Kitaev2001,Ivanov2001,
  Nayak2007, Alicea2010, Stern2013}, Majorana fermions in
con\-densed-matter systems have attracted a great deal of interest.
However, their unambiguous detection in experiments has remained a
difficult task: they are chargeless and -- like the Laughlin
quasiparticles in the fractional quantum Hall effect -- cannot be
extracted from their many-body environment. Elaborate schemes leading
to indirect but conclusive signatures of their presence have been
proposed and partially realized.

Recently, several groups~\cite{Mourik2012,Das2012} have reported the
identification of Majorana bound states in nanowires by observing a
zero-bias peak in tunneling spectroscopy experiments. There is,
however, no consensus regarding the attribution of this result to the
presence of Majorana fermions.
The situation is similar with the experimental
report~\cite{Rokhinson2012} of a 4$\pi$-periodic Josephson effect,
which cannot yet be unambiguously attributed to the presence of
Majorana bound states in the Josephson junction.

There are also many proposals to detect Majorana fermions based on
interferometric structures. They can be divided into two classes. The
first class intends to probe the non-Abelian statistics of Majorana
bound states trapped in vortices of topological superconductors. In
Ref.~\cite{Akhmerov2009,Nilsson2010}
the authors study conductance signatures of vortex tunneling in a
Fabry-P\'erot interferometer.  Another proposal~\cite{Grosfeld2010} is
based on a Mach-Zehnder interferometer constructed from a topological
Josephson junction. In that case, Josephson vortices trapping a
Majorana bound state propagate along the two arms of the
interferometer and give rise to a Josephson-vortex current $I_v$. The
presence of absence of another MBS at the center of the
interferometer, which can be tuned by a magnetic flux, leads to a
striking switching between a vanishing $I_v$ and a nonzero $I_v$. The
roots of this effect are traced back to the non-Abelian exchange
statistics of Majorana bound states.

The goal of the second class of interferometer-based proposals is to
find a signature of Majorana edge states~\cite{Akhmerov2009,Fu2009,Liu2011,struebi2011,bose2011,buettiker2012}.
They propose to use 3D topological insulator heterostructures to build
a Mach-Zehnder interferometer for Majorana edge states contacted by
electronic leads and find conductance signatures.
Reference~\cite{Liu2011} finds a signature of chiral Majorana modes in
the conductance of a Mach-Zehnder interferometer built in a
superconductor--quantum spin Hall--superconductor sandwich.

B\"uttiker {\it et al.}~\cite{buettiker2012} introduced a powerful
scattering matrix approach in terms of Majorana modes that highlights
their special properties and is readily applied to these
interferometric structures.

Motivated by these works, in Ref.~\cite{struebi2011} this setup was
extended to interferometers with two incoming and two outgoing Dirac
fermion channels. Furthermore it was proposed to study noise
correlations in a Hanbury Brown-Twiss (HBT) type interferometer, and
three signatures of the Majorana nature of the channels were
predicted.  First, the average charge current in the outgoing leads
vanishes.  Furthermore, an anomalously large shot noise in the output
ports for a vanishing average current signal is expected. Adding a
quantum point contact (QPC) to the setup, a surprising absence of
partition noise was found which can be traced back to the Majorana
nature of the carriers.

In view of previous successes of studies of higher-order
correlators \cite{Levitov1993,Muzykantskii1994,Belzig2001,boerlin2002}
it is natural to ask~\cite{thesis,beenakker2015} if other Majorana signatures
could be hidden in higher-order correlations. Therefore we will investigate
the full counting statistics in multi-terminal structures containing
Majorana modes to access correlations beyond noise.

In Ref.~\cite{weithofer2014}, the full counting statistics was
calculated for a network of localized Majorana bound states coupled by
tunneling. In this work, in contrast, we will use full counting
statistics to focus on the transport in systems containing
one-dimensional, propagating Majorana modes.

\section{Scattering matrix and full counting statistics (FCS)}
A very general description of quantum transport in a multi-terminal
device is to calculate the full counting statistics (FCS) of the
transfered charges, which contains the full information about the
current and higher-order current-current correlation functions at zero
frequency. Let us denote the probability that $N_1,N_2,\ldots$ charges
are transported into terminal $1,2,\ldots$ during a fixed measurement
time by $P(N_1,N_2,\ldots)$. A quantity which is equivalent but easier
to calculate is the cumulant generating function (CGF), defined by
$\ln \chi(\lambda_1,\lambda_2,\ldots)=\ln \langle
\exp(i\lambda_1N_1+i\lambda_2N_2+\cdots)\rangle$,
where the average denotes the statistical average with the probability
$P$.

FCS helps to understand transport because it allows an identification
of elementary transport events by decomposing the CGF into a sum of
multinomial distributions, where the sum indicates the independence of
the various transport processes in the long-time limit.

\begin{figure}
\centering
  \includegraphics[width=0.5\linewidth]{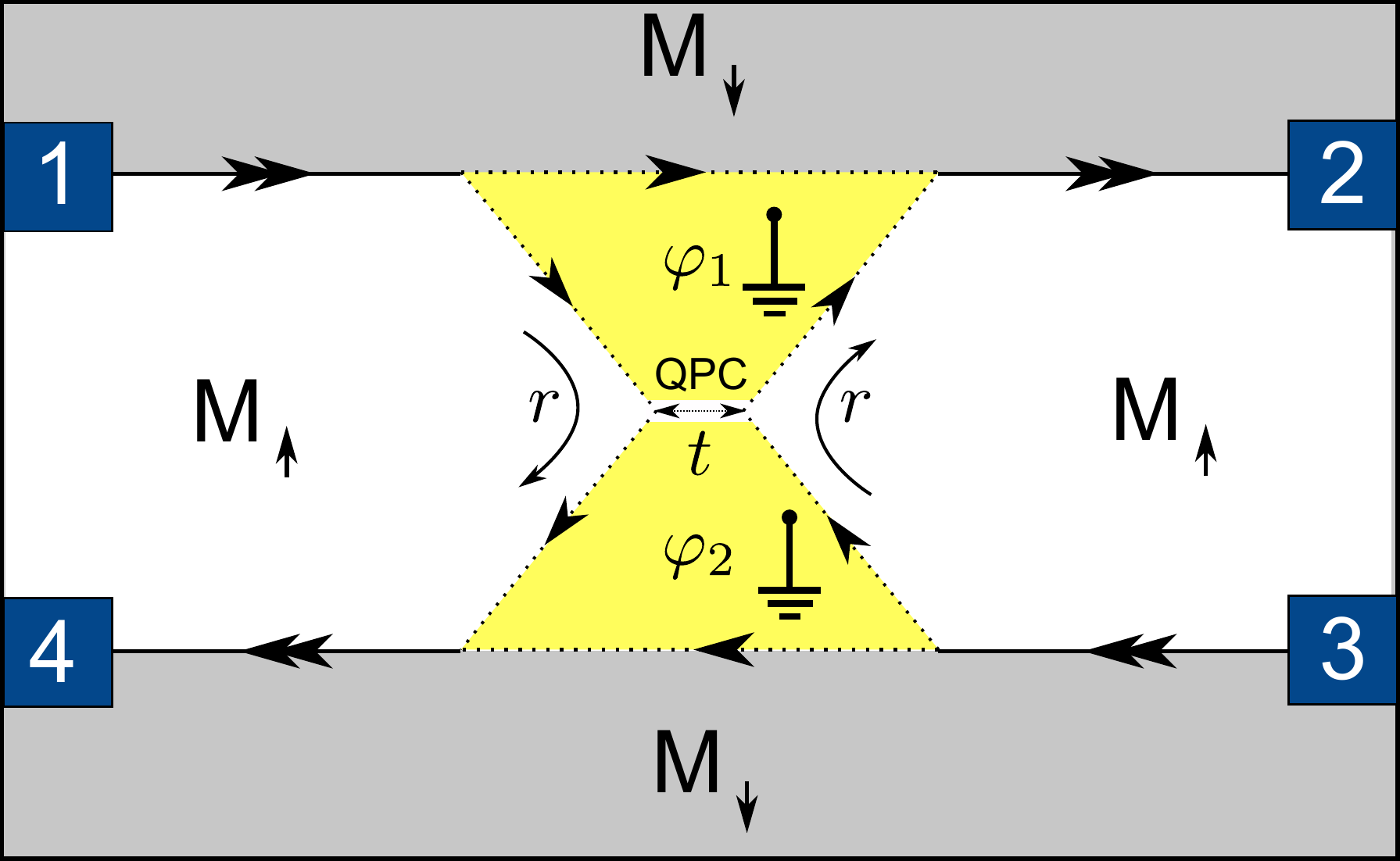}
  \caption{Modified Hanbury Brown-Twiss interferometer, see \cite{struebi2011}.
 Majorana excitations will propagate along the boundaries of the two
    triangular superconducting structures with phases $\varphi_1$,
    $\varphi_2$. An additional short gapped channel appears at
    the domain wall between the two superconducting regions, forming a
    quantum point contact characterized by reflection and transmission
    amplitudes $r$, $t$.
The setup is similar to the one proposed in \cite{Fu2009}.}
  \label{fig:HBTQPC}
\end{figure}

Our goal is to study Majorana interferometers like, e.g., the Hanbury
Brown-Twiss interferometer with the additional QPC, see
Fig.~\ref{fig:HBTQPC} and \cite{struebi2011}.

In the absence of interactions, the FCS can be
obtained from the $4\times4$ scattering matrix $S$ that
relates the outgoing Dirac fermions to the incoming Dirac fermions in
the following way
\begin{align}
\left(\begin{array}{c}\psi_{2,e}\\\psi_{2,h}\\\psi_{4,e}\\\psi_{4,h}\end{array}\right) = S \left(\begin{array}{c}\psi_{1,e}\\\psi_{1,h}\\\psi_{3,e}\\\psi_{3,h}\end{array}\right)\:,
\end{align}
where $1,3$ label the incoming channels, $2,4$ the outgoing channels,
$e$ denotes an electronic mode, and $h$ denotes a hole mode.


The matrix elements of the scattering matrix represent the probability
amplitude of different ``processes''. For instance, the matrix element
$s_{11}$ is the probability amplitude that an incoming electron in
lead 1 goes out as an electron in lead $2$; and $s_{34}$ is the
probability amplitude that an incoming hole in lead 3 goes out as an
electron in lead 4.
The matrix elements already
contain all the interference effects of single-particle states.

{\bf Two-particle processes}: To
obtain the probability amplitudes for two-particle states, we must
take anti-symmetric combinations of products of matrix elements. For
instance, the probability amplitude that two incoming electrons in
leads $1$ and $3$ go out as electrons in leads $2$ and $4$ is given by
$s_{11}s_{33}-s_{13}s_{31}$, i.e., the probability amplitude that the
incoming electron in lead 1 goes out as an electron in lead 2 AND the
incoming electron in lead 3 goes out as an electron in lead 4, MINUS
the probability amplitude that the incoming electron in lead 3 goes
out as an electron in lead 2 AND the incoming electron in lead 1 goes
out as an electron in lead 4. These two processes interfere and lead
to effects such as the two-particle Aharonov-Bohm effect~\cite{Samuelsson2004}.

{\bf Three-particle processes}: It is also possible to have
three-particle processes: an incoming lead populated by both an
electron and a hole, the other incoming lead populated by either an
electron or a hole.

{\bf Four-particle process}: There is only one such process:
all the incoming states and all the outgoing states are totally
filled. This (trivial) process occurs with probability amplitude
$\det S=1$ (or probability $|\det S|^2=1$).

To obtain the full counting statistics we sum up the
contributions of all these coherent processes weighted by the
probability of the input states (occupation of the incoming
leads). For instance, the probability $P_{ij}$ of obtaining
two particles in the outgoing modes $i,j \in \{1, \ldots 4\}$ is
\begin{align}
  P_{ij} = \frac{1}{2} \sum_{k,l = 1}^{4} \left|s_{ik}s_{jl} - s_{il}
    s_{jk}\right|^2 n_k n_l (1-n_a)(1-n_b)\:,
\label{probabilities}
\end{align}
where $k,l$ run over the incoming modes, and $n_k$ is the occupation
of the incoming mode $k$. The indices $a$ and $b$ are the two
remaining incoming modes (such that the set $\{a,b,k,l\} =
\{1,2,3,4\}$). The factor $1/2$ is inserted to avoid double
counting.

A measurement of the current and its correlation function will not
make is possible to determine all entries of the matrix $P_{ij}$
because, e.g., processes with no charge going out at all are not
distinguishable from processes where both an electron and a hole go
out into the same lead. Therefore, the FCS is characterized by only
nine independent probabilities:
\begin{itemize}
\item $P_{00}$, no net charge in the outgoing leads,
\item $P_{\pm0}$, an electron (hole) in the outgoing lead 2,
\item $P_{0\pm}$, an electron (hole) in the outgoing lead 4,
\item $P_{\pm\pm}$, an electron (hole) in both outgoing leads,
\item $P_{\pm\mp}$, an electron (hole) in outgoing lead 2 and a hole
(electron) in outgoing lead 4.
\end{itemize}
All of these probabilities can be conveniently retrieved from the
scattering matrix and the occupation of the leads. The cumulant
generating function is $\ln\chi(\lambda_2,\lambda_4)$, where
$\lambda_i$ is the counting field of outgoing charge in lead $i$, and
\begin{align}
  \chi(\lambda_2,\lambda_4) =  \sum_{s_2, s_4 = -,0,+}
  P_{s_2s_4} e^{ie(s_2\lambda_2 + s_4\lambda_4)} \:.
\label{chi}
\end{align}
Since we treat the leads as free fermion reservoirs the knowledge
of the probabilities allows to directly access the FCS. The cumulant
generating function can be alternatively obtained from the Levitov-Lesovik
determinant formula \cite{Levitov1993,Klich2002}, which can easily be
shown to lead to the same result.

\section{FCS of the Hanbury Brown-Twiss interferometer}
To analyze the full counting statistics of the Hanbury Brown-Twiss
interferometer of Majorana fermions shown in Fig.~\ref{fig:HBTQPC}, we
insert the appropriate expression for the scattering matrix
\cite{struebi2011}
\begin{equation}
\left(\begin{array}{c}
\psi_{2,e}\\\psi_{2,h}\\\psi_{4,e}\\\psi_{4,h}
\end{array}\right) =
\frac{1}{2}\left(\begin{array}{cccc}
\eta_1 - t & \eta_1+t & r & r\\
\eta_1+t & \eta_1-t & -r & -r\\
r & -r & -\eta_2+t & \eta_2+t\\
r & -r & \eta_2+t & -\eta_2+t
\end{array}\right)
\left(\begin{array}{c}
\psi_{1,e}\\\psi_{1,h}\\\psi_{3,e}\\\psi_{3,h}
\end{array}\right).
\label{eq:scatt-mat-hbt-qpc}
\end{equation}
into Eqs.~(\ref{probabilities}) and (\ref{chi}). The probabilities are
related to charge transfer processes of the structure. The probability
$P_{0,0}$ does not contribute to the sum in
Eq.~(\ref{chi}) because it corresponds to processes
where no net charge is transmitted to either lead 2 or 4. The remaining probabilities are given by
\begin{align}
P_{\pm,0}=& \frac{1}{8} - 2 \alpha\beta\gamma\delta \pm \frac1 4  t\ \mathrm{Re}(\eta_1) (\alpha-\beta)(4\gamma\delta+1) +\frac{1}{2} t^2\  (\gamma\delta-\alpha\beta)\: ,\nonumber\\
P_{0,\pm}=& \frac{1}{8} - 2 \alpha\beta\gamma\delta \pm \frac1 4  t\ \mathrm{Re}(\eta_2) (\gamma-\delta)(4\alpha\beta+1) +\frac{1}{2} t^2\  (\alpha\beta-\gamma\delta)\: ,\nonumber\\
P_{\pm,\pm}=& \frac{1}{16} + \alpha\beta\gamma\delta - \frac{1}{8}(\alpha+\beta)(\gamma+\delta)+\frac{1}{8} \mathrm{Re}(\eta_1\eta_2) (\alpha-\beta)(\gamma-\delta)\nonumber\\
&\pm \frac{1}{8} t \left(\mathrm{Re}(\eta_1) (\alpha-\beta)(1-4\gamma\delta)+\mathrm{Re}(\eta_2) (\gamma-\delta)(1-4\alpha\beta)\right)\nonumber\\
&+ \frac{1}{8} t^2 \left(\mathrm{Re}(\eta_1\eta_2^*)(\alpha-\beta)(\gamma-\delta)-(\alpha-\gamma)(\beta-\delta)-(\alpha-\delta)(\beta-\gamma)\right)\: ,\nonumber\\
P_{\pm,\mp}=& \frac{1}{16} + \alpha\beta\gamma\delta - \frac{1}{8}(\alpha+\beta)(\gamma+\delta)-\frac{1}{8} \mathrm{Re}(\eta_1\eta_2) (\alpha-\beta)(\gamma-\delta)\nonumber\\
&\pm \frac{1}{8} t \left(\mathrm{Re}(\eta_1) (\alpha-\beta)(1-4\gamma\delta)-\mathrm{Re}(\eta_2) (\gamma-\delta)(1-4\alpha\beta)\right)\nonumber\\
&- \frac{1}{8} t^2 \left(\mathrm{Re}(\eta_1\eta_2^*)(\alpha-\beta)(\gamma-\delta)+(\alpha-\gamma)(\beta-\delta)+(\alpha-\delta)(\beta-\gamma)\right)\: ,
\label{eq:probFCS}
\end{align}
where we introduced the symmetrized occupation numbers
$\alpha = \frac{1}{2} - n_{1e}$, $\beta = \frac{1}{2} - n_{1h}$,
$\gamma = \frac{1}{2} - n_{3e}$, and $\delta = \frac{1}{2} - n_{3h}$
to simplify the expressions. Since the sum of all probabilities has to be equal to one, one finds
\begin{align}
P_{0,0} =& \frac{1}{4}
  +\frac{1}{2}(\alpha+\beta)(\gamma+\delta)+4\alpha\beta\gamma\delta\nonumber\\
&+\frac{1}{2}
  t^2[(\alpha-\gamma)(\beta-\delta)+(\alpha-\delta)(\beta-\gamma)]\:.
\end{align}

The probabilities $P_{s_2,s_4}$ take into account all the physical
processes that may occur in the interferometer. Processes with
$s_i = \pm1$ correspond to an outgoing electron (hole) in lead i,
while processes with $s_i = 0$ correspond to processes with no
outgoing particles in lead $i$ or to processes with an outgoing
electron \emph{and} hole in lead $i$.

To better understand the probabilities it is
useful to distinguish five classes of processes depending on the
number of incoming particles they involve: the trivial 0-particle
process that contributes solely to $P_{0,0}$; 1-particle processes
that contribute to $P_{0,\pm}$ and $P_{\pm,0}$; 2-particle processes
that contribute to $P_{\pm,\pm}$, $P_{\pm,\mp}$, and $P_{0,0}$;
3-particle processes that contribute to $P_{0,\pm}$ and $P_{\pm,0}$;
and the trivial 4-particle process that also only contributes to
$P_{0,0}$.

\subsection{Results for the cumulants}
To get all cumulants of the outgoing currents, we have to consider the function
\begin{equation}
\frac1h\int_0^\infty dE\ \ln \chi =\frac1h\int_0^\infty dE\ \ln \left(1 + \sum_{s_2,s_4} P_{s_2,s_4} \left(e^{ie(s_2\lambda_2+s_4\lambda_4)}-1\right)\right)\,.
\end{equation}
The cumulants $C(m,n)$ are given by the derivatives of this function at $\lambda_{2,4} = 0$,
\begin{equation}
C(m,n) = \left[(-i\partial_{\lambda_2})^{m}(-i\partial_{\lambda_4})^{n}\frac1h\int_0^\infty dE\ \ln \chi\right]_{\lambda_2=\lambda_4=0}\:.
\end{equation}
As a check we can rederive the results of \cite{struebi2011}. For
example, the average current in lead 2 is given by
\begin{align}
\langle I_2\rangle =& C(1,0) = \frac{e}{h}\int_0^\infty dE\ \sum_{s_2}s_2\sum_{s_4}P_{s_2,s_4}\nonumber\\
=& \frac{e}{h}\int_0^\infty dE\ (P_{++}+P_{+0}+P_{+-}-P_{-+}-P_{-0}-P_{--})\nonumber\\
=& \frac{e}{h}\int_0^\infty dE\ t \mathrm{Re}(\eta_1)v_1\nonumber\\
=& \frac{e}{h}\int_0^\infty dE\ t \mathrm{Re}(\eta_1)(n_{1e}-n_{1h})\,.
\end{align}
making contact to the previously obtained result. Similarly the
current-current cross-correlations read
\begin{align}
S_{24} =& C(1,1) = \frac{e^2}{h}\int_0^\infty dE\
          \sum_{s_2,s_4}s_2s_4P_{s_2,s_4}\nonumber\\
&-\left(\sum_{s_2,s_4}s_2P_{s_2,s_4}\right)\left(\sum_{s_2,s_4}s_4P_{s_2,s_4}\right)\nonumber\\
=& -\frac{e^2 R}{h}\int_0^\infty dE\  (n_{1,e}-n_{1,h})(n_{3,e}-n_{3,h})\mathrm{Re}(\eta_1^*\eta_2)\,
\end{align}
in agreement with the results of \cite{struebi2011}.

\begin{figure}
\begin{center}
	\includegraphics[width=0.5\columnwidth]{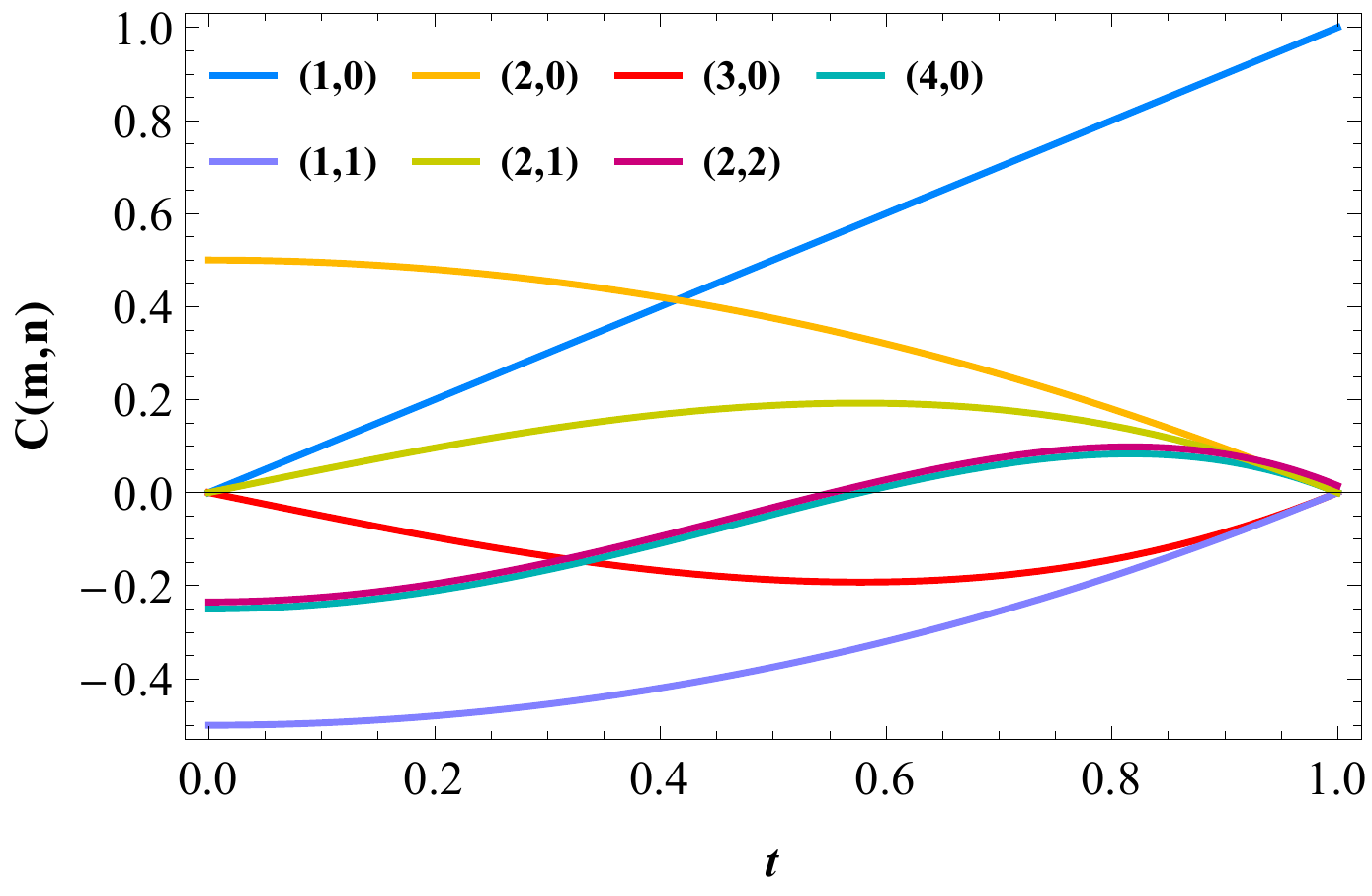}\\
\end{center}
\caption{Cumulants $C(m,n)$ as a function of $t$ (transmission amplitude of the QPC) at zero temperature and for $\eta_1 = 1$, $\eta_2 = -1$. The vertical axis is in arbitrary units and we set $V_1=V_3=e=h=1$.}
\label{fig:cumulants}
\end{figure}

To go further, let us now specify a set of parameters which capture
the most interesting physics. Temperature and interferometer asymmetry
do not bring any new effects, but only smear out certain
quantities. We will thus set $\eta_{1,2} = \pm1$, and $T=0$. Moreover
we will adopt a symmetric voltage configuration where
$V_1 = V_3 = V > 0$ to access the nontrivial multiparticle
processes. A plot of the resulting cumulants is shown in
Fig.~\ref{fig:cumulants}. Two observations can be drawn from this
plot. First, it appears that there is a relation of the form
$|C(m,n)| = |C(m+n,0)|$. Second, the cumulants $C(n,0)$ are connected
to the cumulants of the binomial distribution as discussed in the next
section.

The relation $|C(m,n)| = |C(m+n,0)|$ holds because we treat a special
case : first $P_{0\pm} = P_{\pm0}=0$ and second, for
$\eta_{1,2} =\pm1$, it turns out that either $P_{++} = P_{--} = 0$ or
$P_{+-} = P_{-+} = 0$. It then follows that $S(\lambda_2,\lambda_4)$ is
either a function of $\lambda_2-\lambda_4$ or a function of
$\lambda_2+\lambda_4$. The cumulants of the $k$-th order $C(m,n),\ m+n=k$,
obtained by taking $m$ derivatives of $S$ with respect to $\lambda_2$ and
$n$ derivatives with respect to $\lambda_4$, are thus exactly the same in
the latter case (function of $\lambda_2+\lambda_4$) and only change sign in
the former (function of $\lambda_2-\lambda_4$).

\subsection{Half-charge transfers}
The cumulants $C(n,0)$ are obtained from the cumulant generating function
\begin{equation}
\ln \chi(\lambda) = \ln\left( 1 + P_+ \left(e^{i\lambda}-1\right)+ P_- \left(e^{-i\lambda}-1\right)\right)\: ,
\end{equation}
which follows from the full cumulant generating
function~(\ref{chi}) by summing over the outcomes in
lead 4, and where
$P_\pm = (1\pm\eta_1 t)^2/4$
are the probabilities of transfering an electron or a hole into lead
2. Note that we still assume $T=0, \eta_{1,2}=\pm1, V_1=V_3 = V > 0$.

We may add the side remark that the CGF in the case of $t=0$, viz. a
simple Mach-Zehnder interferometer, corresponds to a trinomial process
with equal probabilities $1/4$ for an electron to electron or to hole
transfer process. This has been noted in \cite{struebi2011}, where the
corresponding predictions for the conductance (it vanishes) and the
noise (is quantized with a Fano factor 1/4) have been
obtained. Recently a detailed analysis of the FCS has confirmed this
prediction \cite{beenakker2015} and related it to the topological
nature of the Majorana mode.

We are now ready to make the connection with the binomial
process. Basically, it turns out that
\begin{align}
\ln \chi(\lambda) =& 2\ln\left(1+\frac{1+\eta_1t}{2}(e^{i\lambda/2}-1)+\frac{1-\eta_1t}{2}(e^{-i\lambda/2}-1)\right)\,.
\label{eq:half-charge}
\end{align}
This means that we can interpret unit-charge transfer processes as two
independent half-charge transfer processes (notice the 2 multiplying
the logarithm, and the factor $\lambda/2$ in the exponents).  In other
words, two independent binomial processes occuring with probability
$p=(1+\eta_1)/2$.

However, we do not expect physical half-charge transport processes
because the existence of charge $e/2$ quasiparticles in this system is
unlikely. We can also write Eq.~(\ref{eq:half-charge}) in the
equivalent form
\begin{align}
\ln \chi(\lambda) =&i\lambda + 2\ln\left(1+\frac{1-\eta_1t}{2}(e^{-i\lambda}-1)\right)\:,
\label{eq:conversion-proc}
\end{align}
which has a more mundane interpretation.
Equation~(\ref{eq:conversion-proc}) represents two independent conversion
processes: two incoming electrons can either go out as electrons with
probability $\sqrt{P_+}$ each, or be converted to a hole with
probability $\sqrt{P_-}$.

Both interpretations, with half-charges or with conversion processes,
are deceptive. In reality, the two incoming electrons are not
independent. However, the factorization of the process itself,
regardless of its interpretation, is of special interest. This is a
potential signature of Majorana fermions. We would like to point out
that this factorization is only valid for a symmetric interferometer,
or at zero bias. This could be potentially attributed to a special
property of the $k=0$ Majorana mode (which is self-adjoint) that the
$k\neq0$ modes do not possess.

Similar half-charge full counting statistics have been found in
other mesoscopic transport situations, e.g. a double-barrier
structure (see Eq.~(38) of \cite{Dejong1996} or Eq.~(3.51) of
\cite{NazarovBlanter}), or a voltage-driven quantum point contact
strongly coupled to a charge qubit \cite{Nazarov2007}.

\section{Factorization of transfer processes}
To shed more light on this question, we will now study this
factorization property for a general scattering matrix.
As discussed before, in some circumstances, it is possible to factorize
the charge transfer processes, such that the cumulant generating
function takes the form
\begin{align}
\ln\chi(\lambda_2,\lambda_4) = 2 \ln \chi'(\lambda_2,\lambda_4)\:,
\end{align}
where the ``half-CGF'' $\ln\chi'$ is expressed in terms of
half-charge transfers
\begin{align}
  \ln\chi'(\lambda_2,\lambda_4) = \ln \sum_{s_2, s_4 = -,0,+}
  P'_{s_2s_4} e^{ie(s_2\lambda_2 + s_4\lambda_4)/2} \:.
\label{chiprime}
\end{align}
Note the factor $1/2$ multiplying the counting fields in the exponent.

To see when this factorization is possible, let us compare $(\chi')
^2$ with $\chi$:
\begin{align}
  (\chi')^2 = \sum_{s_2,s_4,t_2,t_4}
  P'_{s_2s_4}P'_{t_2t_4}e^{ie(\lambda_2(s_2+t_2)/2
    +\lambda_4(s_4+t_4)/2)}
\end{align}
The first conclusion we draw is that we cannot mix 0-charge transfers
with $1/2$-charge transfers in $\chi'$ in the same outgoing
lead. Otherwise, we would still have $1/2$-charge transfers in
$(\chi')^2$. This means in Eq.~(\ref{chiprime}) we must have either $s_{2,4} = \pm1$
or $s_{2,4} = 0$). The case of 0-charge transfer only is not interesting,
so we focus on purely half-charge transfers. We obtain the following
equations for the probabilities
\begin{align}
(P'_{\pm\pm})^2 &= P_{\pm\pm}\:,\nonumber\\
(P'_{\pm\mp})^2 &= P_{\pm\mp}\:,\nonumber\\
2P'_{\pm+}P'_{\pm-} &= P_{\pm0}\:,\nonumber\\
2P'_{+\pm}P'_{-\pm} &= P_{0\pm}\:,\nonumber\\
2P'_{++}P'_{--} + 2P'_{+-}P'_{-+} &= P_{00}\:.
\end{align}
The first two equations fully fix the four probabilities $P'$, the
three remaining equations are thus nontrivial conditions the
probabilities $P$ must satisfy to make the factorization possible.

If we specialize to two-incoming-particle processes, we can refine
the conditions on $P$. First, if we only have
two-incoming-particle processes (this can be tuned by choosing proper
occupations of the incoming leads, such as a symmetric bias and zero
temperature), then $P_{\pm0} = P_{0,\pm} = 0$. The half-charge
transfer process (i.e., the factorization) is then possible if either
\begin{align}
\sqrt{P_{++}} + \sqrt{P_{--}} = 1
\label{fac_prop1}
\quad {\rm or } \quad
\sqrt{P_{+-}} + \sqrt{P_{-+}} = 1\:.
\end{align}

Thus, we are really looking for a property of the
scattering matrix (that is independent of the incoming leads'
occupation) rather than a property of the full counting statistics
themselves. Typically, if two-particle processes and one-particle
processes mix, we have no chance of finding such a factorization of
the FCS.

We now specialize to the case where both electron incoming modes are
fully occupied, and the hole modes fully unoccupied. This leads to the
probabilities
\begin{align}
\sqrt{P_{++}}=|s_{11}s_{33}-s_{13}s_{31}|\:,\nonumber\\
\sqrt{P_{--}}=|s_{21}s_{43}-s_{23}s_{41}|\:,\nonumber\\
\sqrt{P_{+-}}=|s_{11}s_{43}-s_{13}s_{41}|\:,\nonumber\\
\sqrt{P_{-+}}=|s_{21}s_{33}-s_{23}s_{31}|\:.
\label{fac_prop_S}
\end{align}
Thus, checking the factorization property given in
Eq.~(\ref{fac_prop1}) has been reduced to a
condition on the scattering matrix.
The strategy will now be to investigate interferometers of increasing
complexity by writing down their scattering matrices
and checking whether Eq.~(\ref{fac_prop1}) is
fulfilled, i.e., whether the factorization property holds.

\subsection{Double point-contact geometry}
\begin{figure}
\includegraphics[width = 0.45\linewidth]{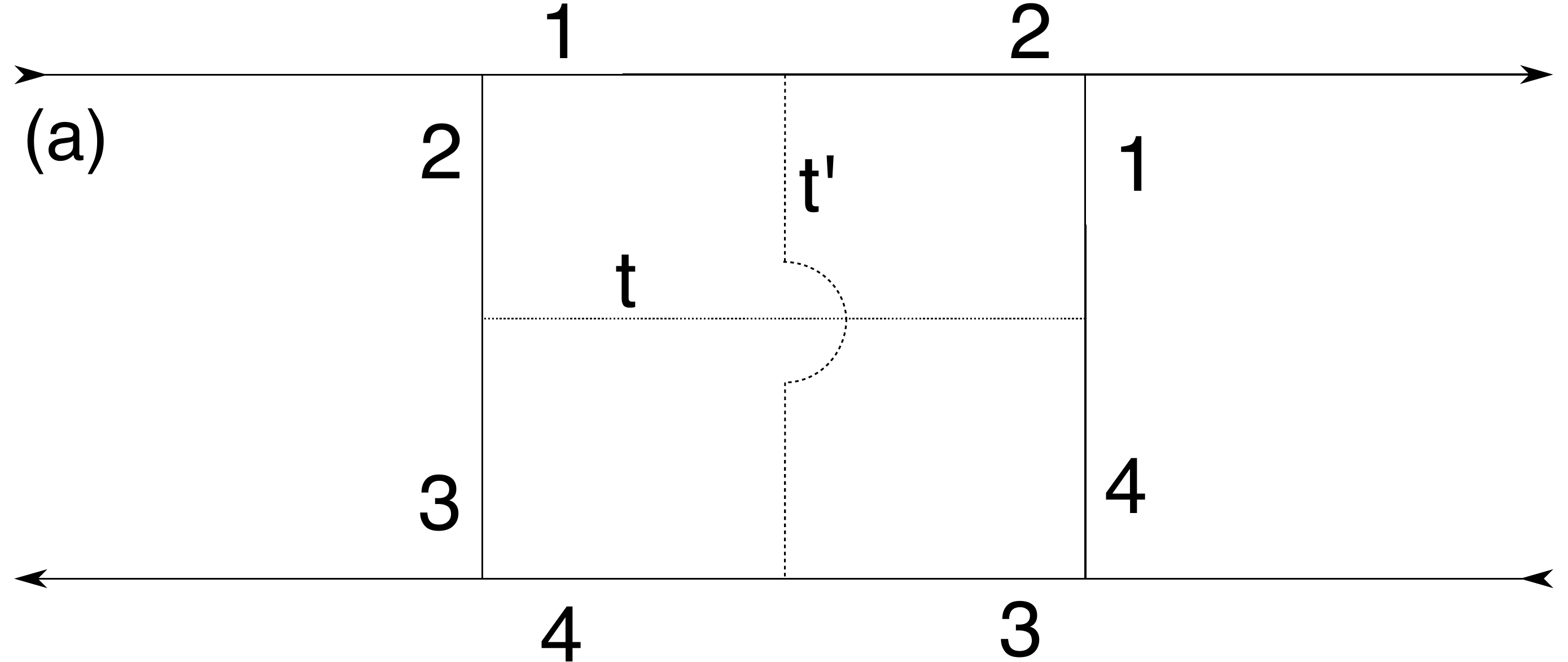}
\qquad\includegraphics[width = 0.45\linewidth]{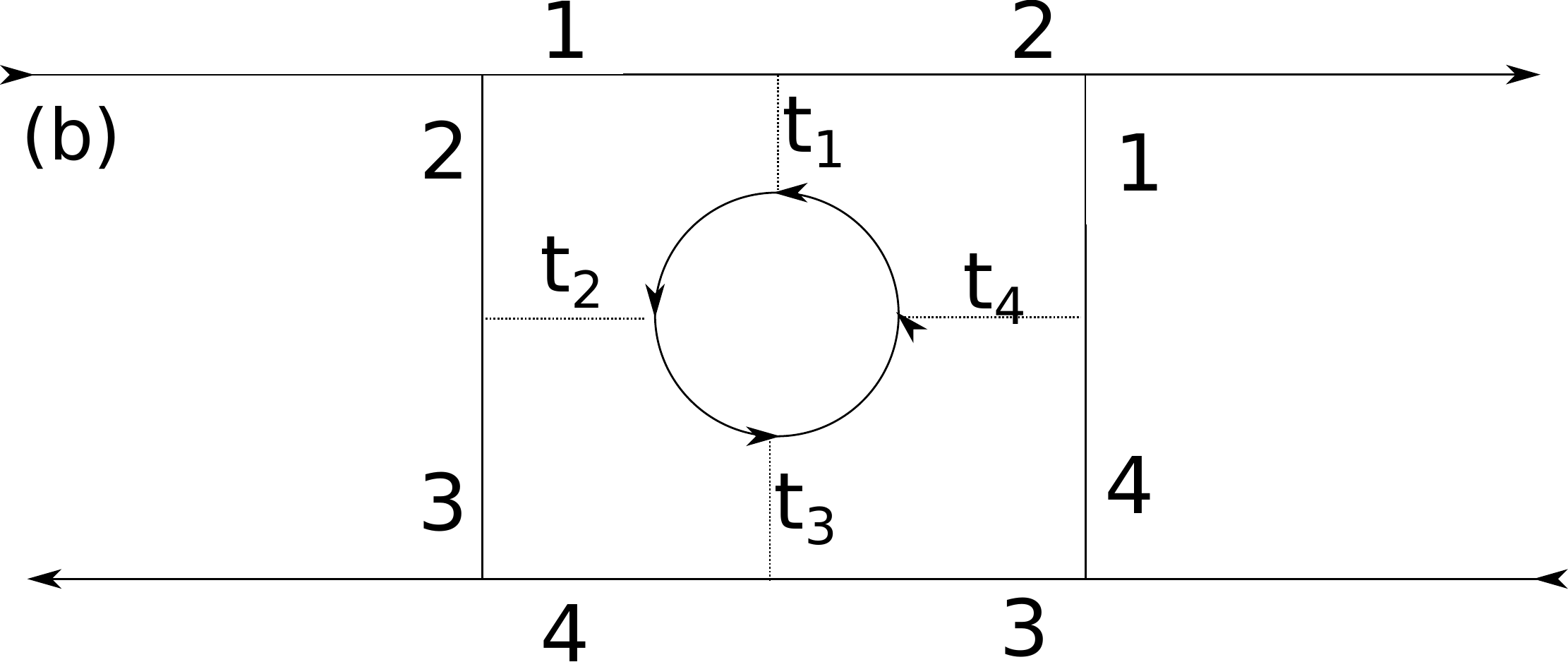}
\center{\includegraphics[width = 0.45\linewidth]{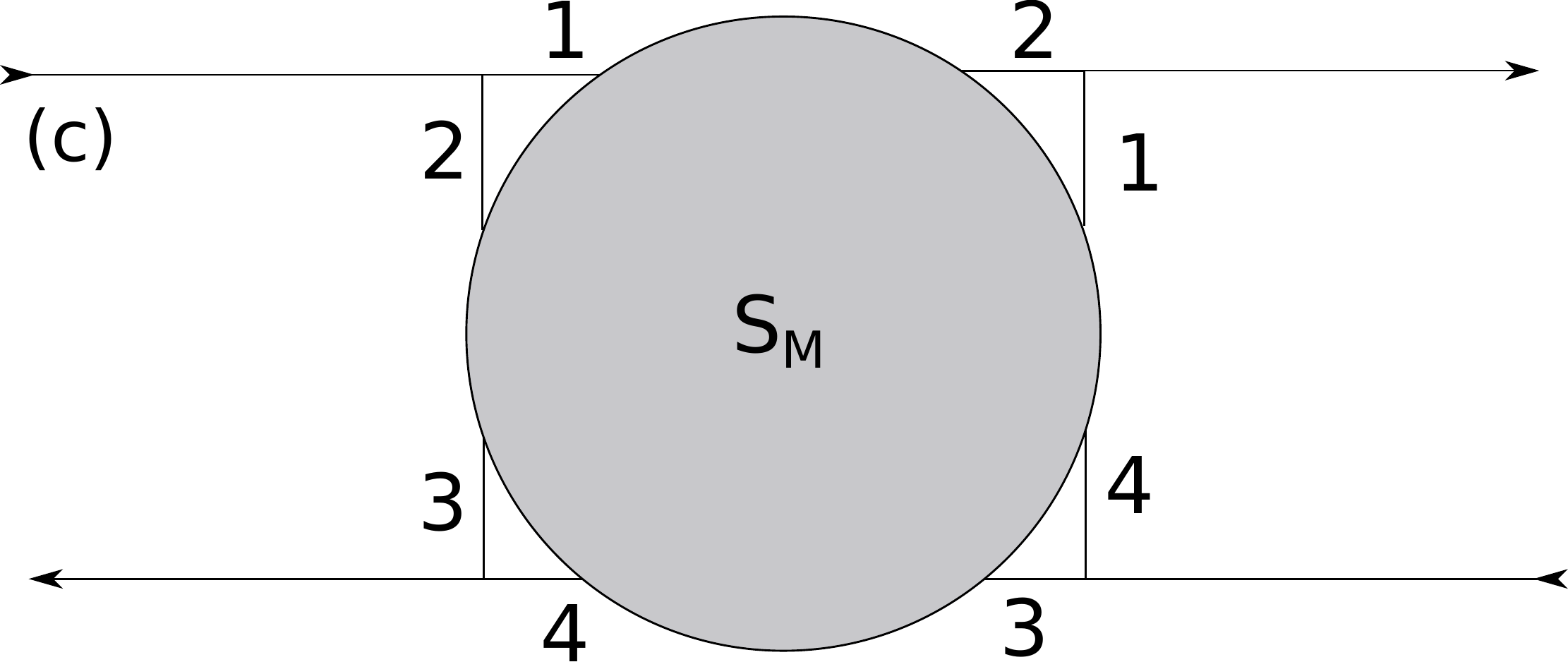}}
\caption{
(a) Double point-contact geometry.
(b) Central island geometry.
(c) General scattering matrix. The numbers $1,2,3,4$ in the top-left and botton-right corners label the four incoming chiral Majorana modes, whereas the numbers in the top-right and bottom-left corners label the four outgoing Majorana modes. Incoming and outgoing Majorana modes are related by the scattering matrix $S_M$.}
\label{label_fig1}
\end{figure}

We first consider the double point-contact interferometer shown in
Fig.~\ref{label_fig1}a where the labels for the incoming
and outgoing Majorana modes are defined.

The conversion of Dirac to Majorana modes is achieved by \cite{Akhmerov2009}
\begin{align}
S_{c} = \frac1{\sqrt2}\left(\begin{array}{cccc}\eta_A & \eta_A & 0 & 0\\ i & -i & 0 & 0\\  0 & 0 & -\eta_C & -\eta_C \\ 0 & 0 & i & -i\end{array}\right)\:,
\end{align}
see~\ref{interferometer_basics}. Here $\eta_A,\eta_C = \pm 1$
depend on the parity of the number of vortices. The minus sign in
front of $\eta_C$ takes care of the $\pi$ Berry phase obtained when a
Majorana fermion makes a $2\pi$ rotation. The back-conversion is
similar,
\begin{align}
S_{bc} = \frac1{\sqrt2}\left(\begin{array}{cccc}-i & \eta_B & 0 & 0\\ i & \eta_B & 0 & 0\\  0 & 0 & -i & \eta_D \\ 0 & 0 & i & \eta_D \end{array}\right)\:,
\end{align}
The tunneling of Majorana fermions is described by the $SO(4)$ matrix
\begin{align}
S_{M} = \left(\begin{array}{cccc} 0 & -t & 0 & r\\ r' & 0 & t' & 0\\  0 & r & 0 & t \\ -t' & 0 & r' & 0 \end{array}\right)\:.
\end{align}

The total scattering matrix is given by the product of these three
scattering matrices
\begin{align}
S &= S_{bc} S_M S_c
\end{align}
which is real.
Using Eq.~(\ref{fac_prop_S}), one obtains
\begin{align}
\sqrt{P_{++}}+\sqrt{P_{--}} = \frac{1+\eta}{2}\:,\\
\sqrt{P_{+-}}+\sqrt{P_{-+}} = \frac{1-\eta}{2}\:,
\end{align}
i.e., the sums of the probability amplitudes are seen to fulfill
Eq.~(\ref{fac_prop1}).  Thus, factorization holds, and
the two-incoming-electron processes can be written as two equal and
independent processes.

\subsection{Central Majorana island geometry}

We define the matrix
\begin{align}
X = \left(\begin{array}{cccc}0 & 0 & 0 & 1\\1 & 0& 0 & 0\\0 & 1 & 0 & 0\\0 & 0 & 1 & 0\end{array}\right)\:,
\end{align}
as well as $R=\mathrm{diag}(r_1,r_2,r_3,r_4)$,
and $T=\mathrm{diag}(t_1,t_2,t_3,t_4)$.
Using $(XR)^4 = r_1r_2r_3r_4 \equiv r$ is
proportional to the identity matrix, we find
\begin{align}
S_M = XR -\frac1{1-r} XT (1+XR + (XR)^2+(XR)^3)XT \:.
\end{align}
The relevant matrix elements of the total scattering matrix
$S = S_{bc}S_MS_c$ read
\begin{align}
s_{11} &= \frac{1}{2}\left(r_1-\frac{t_1r_2r_3r_4t_1+t_4r_1t_2-i(t_4t_1-t_1t_2)}{1-r}\right)\:,\nonumber\\
s_{33} &= \frac{1}{2}\left(r_3-\frac{t_3r_4r_1r_2t_3+t_2r_3t_4-i(t_2t_3-t_3t_4)}{1-r}\right)\:,\nonumber\\
s_{13} &= \frac{1}{2}\left(r_4-\frac{t_4r_1r_2r_3t_4+t_1r_2t_3-i(t_4r_1r_2t_3-t_1r_2r_3t_4)}{1-r}\right)\:,\nonumber\\
s_{31} &= \frac{1}{2}\left(r_2-\frac{t_2r_3r_4r_1t_2+t_3r_4t_1-i(t_2r_3r_4t_1-t_3r_4r_1t_2)}{1-r}\right)\:,\nonumber\\
s_{21} &= \frac{1}{2}\left(r_1-\frac{t_1r_2r_3r_4t_1-t_4r_1t_2+i(t_4t_1+t_1t_2)}{1-r}\right)\:,\nonumber\\
s_{43} &= \frac{1}{2}\left(r_3-\frac{t_3r_4r_1r_2t_3-t_2r_3t_4+i(t_2t_3+t_3t_4)}{1-r}\right)\:,\nonumber\\
s_{23} &= \frac{1}{2}\left(-r_4-\frac{-t_4r_1r_2r_3t_4+t_1r_2t_3+i(t_4r_1r_2t_3+t_1r_2r_3t_4)}{1-r}\right)\:,\nonumber\\
s_{41} &= \frac{1}{2}\left(-r_2-\frac{-t_2r_3r_4r_1t_2+t_3r_4t_1+i(t_2r_3r_4t_1+t_3r_4r_1t_2)}{1-r}\right)\:.
\label{S_island}
\end{align}
If only one QPC has a non-vanishing transmission, the
scattering matrix reduces to the HBT case. For $t_2 \equiv t$, $t_4 =
1$, and $t_1=t_3=0$, we get back the setup HBT with one QPC. For the
case where all QPC are equal, $t_i \equiv t$,
the factorization also holds. The same is true for $t_1=t_4\equiv t$ and
$t_2=t_3 = 0$ although this situation is less symmetric.

We were unable to prove the factorization property for general values
$t_i$ analytically.
We therefore used a numerical approach: in the spirit of a Monte-Carlo
calculation, we generated many scattering matrices by using random
values of the transmission amplitudes $t_i$. The factorization was
then confirmed by checking that
$\sqrt{P_{++}} + \sqrt{P_{--}} =
|s_{11}s_{33}-s_{13}s_{31}|+|s_{21}s_{43}-s_{23}s_{41}| = 0 \quad
\mathrm{or} \quad 1$.
This is indeed the case, i.e., the factorization property holds for
the island geometry as well.

\subsection{Most general scattering matrix}


We now consider the most general case: a general $SO(4)$ matrix
accompanied by the appropriate Dirac to Majorana conversion /
back-conversion, see Fig.~\ref{label_fig1}c.
The conversion from Dirac to Majorana channels reads
\begin{align}
S_{c} = \frac1{\sqrt2}\left(\begin{array}{cccc}1 & 1 & 0 & 0\\ i & -i & 0 & 0\\  0 & 0 & -1 & -1 \\ 0 & 0 & i & -i\end{array}\right)\:;
\end{align}
the back-conversion is similar
\begin{align}
S_{bc} = \frac1{\sqrt2}\left(\begin{array}{cccc}-i & 1 & 0 & 0\\ i & 1 & 0 & 0\\  0 & 0 & -i & 1 \\ 0 & 0 & i & 1 \end{array}\right)\:.
\end{align}
The total scattering matrix is
\begin{align}
S = S_{bc}S_MS_{c}\:,
\label{S_general}
\end{align}
where $S_M\in SO(4)$. Again, we demonstrated the factorization property numerically by
generating a set of random $SO(4)$ matrices
and constructing the corresponding scattering matrix $S$ defined by
Eq.~(\ref{S_general}).  The factorization property
(\ref{fac_prop1}) was found to hold even in this most
general case.

\subsection{More than four terminals}

Finally, let us discuss how the peculiar properties of the CGF can be
generalized to setups with more than four terminals. In the case of
six terminals, for instance, a general scattering matrix between
incoming and outgoing Dirac fermions can be expressed as
\begin{align}
S = S_{bc}S_MS_{c}\:,
\end{align}
where $S_M$ is a general $SO(6)$ matrix, and $S_{c}$ and $S_{bc}$
denote the $6 \times 6$ conversion and back-conversion matrices. If
particles enter via the leads $1,3,5$ and exit via the leads $2,4,6$,
then the most general CGF is a generalization of Eq.~(\ref{chi}),
\begin{align}
  \chi(\lambda_2,\lambda_4,\lambda_6) =  \sum_{s_2, s_4, s_6 = -,0,+}
  P_{s_2s_4s_6} e^{i(s_2\lambda_2 + s_4\lambda_4 + s_6 \lambda_6)} \:.
\end{align}
If we focus again on the limit of equal positive bias voltages
$V_1 = V_3 = V_5 > 0$, then all incoming electron modes are occupied
and all incoming hole modes are empty. In that case, in analogy to
Eq.~(\ref{fac_prop_S}), one finds, e.g., for the probability $P_{+++}$
of finding particles in all outgoing modes and the probability
$P_{++-}$ to find particles exiting in leads $2$ and $4$ and a hole
exiting via lead 6, \renewcommand{\P}{\mathcal{P}}
\begin{align}
 \sqrt{P_{+++}} &= \sum_{\P \in S_{135}} |(-1)^{\text{sgn}(\P)} s_{1,\P1} s_{3,\P3} s_{5,\P5} |, \notag \\
 \sqrt{P_{++-}} &= \sum_{\P \in S_{136}} |(-1)^{\text{sgn}(\P)} s_{1,\P1} s_{3,\P3} s_{5,\P6} |.
\end{align}
where the summation is over the six permutations of the respective
set. With our choice of phases in the conversion matrices, one finds
that for an arbitrary $SO(6)$ matrix $S_M$ only processes with
outgoing charge $3$ and $-1$ have a nonzero probability,
\begin{align}
 P_{+++} + P_{+--} + P_{-+-} + P_{--+} &= 1, \notag \\
 P_{++-} = P_{+-+} = P_{-++} = P_{---} &= 0.
\end{align}
Using the first line to eliminate $P_{+++}$, one finds that the CGF is given by a generalization of Eq.~(\ref{eq:conversion-proc}),
\begin{align}
& \ln \chi(\lambda_2,\lambda_4,\lambda_6) = i (\lambda_2 + \lambda_4 + \lambda_6)
 + \ln \bigg[ 1 +
  P_{+--} (e^{-2i(\lambda_4 + \lambda_6)} - 1) \notag \\+
  &P_{-+-} (e^{-2i(\lambda_2 + \lambda_6)} - 1) +
  P_{--+} (e^{-2i(\lambda_2 + \lambda_4)} - 1) \bigg]
\end{align}
Evidently, for a general scattering matrix the six-terminal result
contains more free parameters than the four-terminal case. This means
that it is no longer possible to factorize the FCS in general.

However, one still finds that only a very limited number of scattering
processes are possible: either two of the incoming eletrons are
transmitted as holes (with probabilities $P_{+--}$, $P_{-+-}$ and
$P_{--+}$) or all electrons are transmitted as electrons.

\section{Discussion/Conclusion}
To summarize, we have calculated and analyzed the full counting
statistics of a modified Hanbury Brown-Twiss interferometer for chiral
Majorana fermions which contain information about higher-order current
correlation and generalize the results presented in
\cite{struebi2011}.  Most of the calculations in this paper are valid
only at zero energy, at which particle-hole symmetry enforces
important constraints on the scattering matrices. Some of the obtained
results are expected to remain valid at finite, but low,
energies. However, we also do not keep track of dynamical phases which
are important even at low energies.

The full counting statistics calculated in this paper exhibit an
interesting factorization property that points towards the
interpretation of a unit-charge transfer process as two independent
half-charge transfer processes. We checked the factorization property
for increasingly more general scattering matrices and confirmed
numerically that it holds for the most general $SO(4)$ matrix at
specific configurations of voltage bias and at zero temperature.  It
is tantalizing to interpret this property as a signature of Majorana
fermions, however, since we do not expect physical 1/2 charges to be
transferred, its interpretation remains open at present.

\section{Acknowledgment}
GS and CB acknowledge financial support by the the Swiss SNF and the
NCCR Quantum Science and Technology.  WB was financially supported by
the DFG through SFB 767 and BE 3803/5-1.  TLS acknowledges support by
National Research Fund, Luxembourg (ATTRACT 7556175).

\begin{appendix}

\section{Majorana interferometers basics}
\label{interferometer_basics}
In this appendix, we list some of the basic building blocks needed to
study concrete interferometric structures, compute their
scattering matrix, and check whether the FCS factorization holds.

\subsection{Dirac to Majorana converter}
This is the building block to connect Majorana fermions to external Dirac
fermions~\cite{Akhmerov2009,Fu2009}. The scattering matrix that relates chiral Dirac fermions to chiral
Majorana fermions at T-junctions is fully determined by particle-hole
symmetry and unitarity. Its form is given by
\begin{align}
  S_\mathrm{conv} = \frac1{\sqrt2} \left(\begin{array}{cc} 1 & 1\\i &
      -i\end{array}\right)\:,
\end{align}
such that
\begin{align}
  \left(\begin{array}{c} \gamma_1\\\gamma_2\end{array}\right) =
  S_\mathrm{conv} \left( \begin{array}{c}
      \psi_e\\\psi_h \end{array}\right) \:.
\end{align}
The only remaining freedom is a physically unimportant phase factor
that corresponds to gauge transformations
\begin{align}
  S_\mathrm{conv} \longrightarrow S_\mathrm{conv}
  \left(\begin{array}{cc}e^{i\alpha} &0\\0 &
      e^{-i\alpha}\end{array}\right)\:.
\end{align}
As long as the Dirac fermions in the external channels do not
interfere with one another, these factors are irrelevant.

The back conversion of Majorana to Dirac fermions is described by the
time-reversed scattering matrix
\begin{align}
  S_\mathrm{back-conv} = \frac1{\sqrt2} \left(\begin{array}{cc} 1 & i\\1
      & -i\end{array}\right)\:.
\end{align}

\subsection{Majorana fermion point contact}
Because of particle-hole symmetry, the scattering matrix that
relates chiral Majorana fermions before and after a quantum point
contact (QPC), or any form of tunneling, must be real. The scattering matrix
is therefore a rotation matrix that can be parametrized by a
single angle. We use $r,t$ as the (real) reflection and transmission
amplitudes of the point contacts to write
\begin{align}
  \left(\begin{array}{c} \gamma_1'\\\gamma_2'\end{array}\right) =
  \left(\begin{array}{cc} r & t\\-t & r\end{array}\right)
  \left(\begin{array}{c} \gamma_1\\\gamma_2\end{array}\right)\:.
\end{align}

Note that tunneling between a pair of Majorana channels directly after
the conversion from Dirac fermions can be described by the product of
two scattering matrices
\begin{align}
S_\mathrm{tot} = S_\mathrm{tunnel} S_\mathrm{conv}
& = \frac1{\sqrt2} \left(\begin{array}{cc} r+it & r-it\\-t+ir &
    -t-ir\end{array}\right)\nonumber\\
& = \frac1{\sqrt2} \left(\begin{array}{cc} e^{i\alpha} & e^{-i\alpha}\\ie^{i\alpha} & -ie^{-i\alpha}\end{array}\right)\:,
\end{align}
where we defined $r+it = e^{i\alpha}$. We can write this as
\begin{align}
  S_\mathrm{tot} = S_\mathrm{conv} \left(\begin{array}{cc}e^{i\alpha}
      &0\\0 & e^{-i\alpha}\end{array}\right)\:.
\end{align}
The effect of tunneling is therefore equivalent to a redefinition of
the phases of the incoming Dirac modes and can be disregarded.

\end{appendix}

\end{document}